# Network Enhancement: a general method to denoise weighted biological networks

Bo Wang,[1,*] Armin Pourshafeie,[2,*] Marinka Zitnik,[1,*] Junjie Zhu,[3] Carlos D. Bustamante,[4,5] Serafim Batzoglou,[1,‡,#] and Jure Leskovec[1,5,#]

[1] Department of Computer Science, Stanford University, Stanford, CA, USA
[2] Department of Physics, Stanford University, Stanford, CA, USA
[3] Department of Electrical Engineering, Stanford University, Stanford, CA, USA
[4] Department of Biomedical Data Science, Stanford University, Stanford, CA, USA
[5] Chan Zuckerberg Biohub, San Francisco, CA, USA
[‡] Currently at Illumina Inc.
[*] These authors contributed equally.
[#] Corresponding authors: serafim@cs.stanford.edu, jure@cs.stanford.edu

**Networks are ubiquitous in biology where they encode connectivity patterns at all scales of organization, from molecular to the biome. However, biological networks are noisy due to the limitations of measurement technology and inherent natural variation, which can hamper discovery of network patterns and dynamics. We propose Network Enhancement (NE), a method for improving the signal-to-noise ratio of undirected, weighted networks. NE uses a doubly stochastic matrix operator that induces sparsity and provides a closed-form solution that increases spectral eigengap of the input network. As a result, NE removes weak edges, enhances real connections, and leads to better downstream performance. Experiments show that NE improves gene function prediction by denoising tissue-specific interaction networks, alleviates interpretation of noisy Hi-C contact maps from the human genome, and boosts fine-grained identification accuracy of species. Our results indicate that NE is widely applicable for denoising biological networks.**

Networks provide an elegant abstraction for expressing fine-grained connectivity and dynamics of interactions in complex biological systems [1]. In this representation, the nodes indicate the components of the system. These nodes are often connected by non-negative, (weighted-)edges which indicate the similarity between two components. For example, in protein-protein interaction (PPI) networks, weighted edges capture the strength of physical interactions between proteins and can be leveraged to detect functional modules [2]. However, accurate experimental quantification of interaction strength is challenging [3,4]. Technical and biological noise can lead to superficially strong edges, implying spurious interactions; conversely, dubiously weak edges can hide real, biologically important connections [4–6]. Furthermore, corruption of experimentally derived networks by noise can alter the entire structure of the network by modifying the strength of edges within and amongst underlying biological pathways. These modifications adversely impact the performance of downstream analysis [7]. The challenge of noisy interaction measurements is not unique to PPI networks and plagues many different types of biological networks, such as Hi-C [8] and cell-cell interaction networks [9].

To overcome this challenge, computational approaches have been proposed for denoising networks. These methods operate by replacing the original edge weights with weights obtained based on a diffusion defined on the network [10,11]. However, these methods are often not tested on different types of networks [11], rely on heuristics without providing explanations for why these approaches work, and lack mathematical understanding of the properties of the denoised networks [10,11]. Consequently, these methods may not be effective on new applications derived from emerging experimental biotechnology.

Here, we introduce Network Enhancement (NE), a diffusion-based algorithm for network denoising that does not require supervision or prior knowledge. NE takes as input a noisy, undirected, weighted network and outputs a network on the same set of nodes but with a new set of edge weights (Figure ). The main crux of NE is the observation that nodes connected through paths with high-weight edges are more likely to have a direct, high-weight edge between them [12,13]. Following this intuition, we define a diffusion process that uses random walks of length three or less and a form of regularized information flow to denoise the input network (Figure A and **Methods**). Intuitively, this diffusion generates a network in which nodes with strong similarity/interactions are connected by high-weight edges while nodes with weak similarity/interactions are connected by low-weight edges (Figure B). Mathematically, this means that eigenvectors associated with the input network are preserved while the eigengap is increased. In particular, NE denoises the input by down-weighting small eigenvalues more aggressively than large eigenvalues. This re-weighting is advantageous when the noise is spread in the eigen-directions corresponding to small eigenvalues [14]. Furthermore, the increased eigengap of the enhanced network is a highly appealing property as it leads to accurate detection of modules/clusters [15,16] and allows for higher-order network analysis [12]. Moreover, NE has an efficient and easy to implement closed-form solution for the diffusion process, and provides mathematical guarantees for this converged solution. (Figure B and **Methods**).

## Results

We have applied NE to three challenging yet important problems in network biology. In each experiment, we evaluate the network denoised by NE against the same network denoised by alternative methods: network deconvolution (ND) [10] and diffusion state distance (DSD) [11]. For completeness, we also compare our results to a network reconstructed from features learned by Mashup (MU) [17]. All three of these methods use a diffusion process as a fundamental step in their algorithms and have a closed-form solution at convergence. ND solves an inverse diffusion process to remove the transitive edges, and DSD uses a diffusion-based distance to transform the network. While ND and DSD are denoising algorithms, Mashup is a feature learning algorithm that learns low-dimensional representations for nodes based on their steady-state topological positions in the network. This representation can be used as input to any subsequent prediction model. In particular, a denoised network can be constructed by computing a similarity measure using MU's output features [17].

**NE improves human tissue networks for gene function prediction**

Networks play a critical role in capturing molecular aspects of precision medicine, particularly those related to gene function and functional implications of gene mutation [18,19]. We test the utility of our denoising algorithm in improving gene interaction networks from 22 human tissues assembled by Greene *et al.* [20]. These networks capture gene interactions that are specific to human tissues and cell lineages ranging from B lymphocyte to skeletal muscle and the whole brain [20,21]. We predict cellular functions of genes specialized in different tissues based on the networks obtained from different denoising algorithms.

Given a tissue and the associated tissue-specific gene interaction network, we first denoise the network and then use a network-based algorithm on the denoised edge weights to predict gene functions in that tissue. We use standard weighted random walks with restarts to propagate gene-function associations from training nodes to the rest of the network [22]. We define a weighted random walk starting from nodes representing known genes associated with a given function. At each time step, the walk moves from the current node to a neighboring node selected with a probability that depends on the edge weights, and has a small probability of returning to the initial nodes [22]. The algorithm scores each gene according to its visitation probability by the random walk. Node scores returned by the algorithm are then used to predict gene-function associations for genes in the test set. Predictions are evaluated against experimentally validated gene-function associations using a leave-one-out cross-validation strategy.

When averaged over the four denoising algorithms and the 22 human tissues, the gene function prediction improved by 12.0% after denoising. Furthermore, we observed that all denoising algorithms improved the average prediction performance (Figure A and **Supplementary Note 1**). These findings motivate the use of denoised networks over original (raw) biological networks for downstream predictive analytics. We further observed that gene function prediction performed consistently better in combination with networks revised by NE than in combination with networks revised by other algorithms. On average, NE outperformed networks reconstructed by ND, DSD and MU by 12.3%. In particular, NE resulted in an average a 5.1% performance gain over the second best-performing denoised network (constructed by MU). Following Greene *et al.* [20], we further validated our network enhancement approach by examining each enhanced tissue network in turn and evaluating how well relevant tissue-specific gene functions are connected in the network. The expectation is that function-associated genes tend to interact more frequently in tissues in which the function is active than in other non-relevant tissues [20]. As a result, relevant functions are expected to be more tightly connected in the tissue network than functions specific to other tissues. For each NE-enhanced tissue network, we ranked all functions by the edge density of function-associated tissue subnetworks and examined top-ranked functions. In the NE-enhanced blood plasma network, we found that functions with the highest edge density were blood coagulation, fibrin clot formation, and negative regulation of very-low-density lipoprotein particle remodeling, all these functions are specific to blood plasma tissue (Figure B). This finding suggest that tissue subnetworks associated with relevant functions tend to be more connected in the tissue network than subnetworks of non-tissue-specific functions. The most connected functions in the NE-enhanced brain network were brain morphogenesis and forebrain regionalization, which are both specific to brain tissue (Figure B). Examining edge density-based rankings of gene functions across 22 tissue networks, we found relevant functions consistently placed at or near the top of the rankings, further indicating that NE can improve signal-to-noise ratio of tissue networks.

**NE improves Hi-C interaction networks for domain identification**

The recent discovery of numerous cis-regulatory elements away from their target genes emphasizes the deep impact of 3D structure of DNA on cell regulation and reproduction [23–25]. Chromosome conformation capture (3C) based technologies [25] provide experimental approaches for understanding the chromatin interactions within DNA. Hi-C is a 3C-based technology that allows measurement of pairwise chromatin interaction frequencies within a cell population [8,25]. The Hi-C reads are grouped into bins based on the genetic region they map to. The bin size determines the measurement resolution.

Hi-C read data can be thought of as a network where genomic regions are nodes and the normalized count of reads mapping to two regions are the weighted edges. Network community detection algorithms can be used on this Hi-C derived network to identify clusters of regions that are close in 3D genomic structure [26]. The detected megabase-scale communities correspond to regions known as topological associating domains (TADs) and represent chromatin interaction neighborhoods [26]. TADs tend to be enriched for regulatory features [27,28] and are hypothesized to specify elementary regulatory micro-environment. Therefore, detection of these domains can be important for analysis and interpretation of Hi-C data. The limited number of Hi-C reads, hierarchical structure of TADs and other technological challenges lead to noisy Hi-C networks, and hamper accurate detection of TADs [25].

To investigate the ability of NE in improving TAD detection, we apply NE to a Hi-C dataset and analyze the performance of a standard domain identification pipeline with and without a network denoising step. For this experiment, we used 1kb and 5kb resolution Hi-C data from all autosomes of the GM12878 cell line [8]. Since true gold-standards for TAD regions are lacking, a synthetic dataset was created by stitching together non-overlapping clusters detected in the original work [8] as a result, the clusters stitched together can be used as a good proxy for the true cluster (more details in **Supplementary Note 1**.) Figure A shows a heatmap of the raw Hi-C data for a portion of chromosome 16.

We applied two, off the shelf, community detection methods (Louvian [29] and MSCD [30]) to each Hi-C network and compared the quality of the detected TADs with or without network denoising. Visual inspection of the Hi-C contact matrix before and after the Hi-C network is denoised using NE reveals an enhancement of edges within each community and sharper boundaries between communities (Figure A). This improvement is particularly clear for the 5kb resolution data, where communities that were visually undetectable in the raw data become clear after denoising with NE. To quantify this enhancement, the communities obtained from raw networks and networks enhanced by NE or other denoising methods were compared to the true cluster assignments. We used normalized mutual information (NMI, **Supplementary Note 2**) as a measure of shared information between the detected communities and the true clusters. NMI ranges between 0 to 1, where a higher value indicates higher concordance and 1 indicates an exact match between the detected communities and the true clusters. The results across 22 autosomes indicate that while denoising can improve the detection of communities, not all denoising algorithms succeed in this task (Figure B). For both resolutions considered, NE performs the best with an average NMI of 0.92 for 1kb resolution and 0.94 for 5kb resolution, MU (the second best performing method) achieves an average NMI of 0.85 and 0.84, respectively while ND and DSD achieve lower average NMI than the raw data which has NMI of 0.81 and 0.67, respectively. Furthermore, we note that the performance of NE and MU remains high as the resolution decreases from 1kb to 5kb, in contrast the ability of the other pipelines in detecting the correct communities diminishes. While MU maintains a good average performance at 5kb resolution, the standard deviation of NMI values after denoising with MU increases from 0.037 in 1kb data to 0.054 in 5kb data due to relatively poor performances on a few chromosomes. On the other hand, the NMI values for data denoised with NE maintain a similar spread at both resolutions (standard deviation 0.033 and 0.031, respectively). The better average NMI and smaller spread indicates that NE can reliably enhance the network and improve TAD detection.

**NE improves similarity network for fine-grained species identification**

Fine-grained species identification from images concerns querying objects within the same subordinate category. Traditional image retrieval works on high-level categories (*e.g.*, finding all butterflies instead of cats in a database given a query of a butterfly), while fine-grained image retrieval aims to distinguish categories with subtle differences (*e.g.*, monarch butterfly versus peacock butterfly). One major obstacle in fine-grained species identification is the high similarity between subordinate categories. On one hand, two subordinate categories share similar shapes and carry subtle color difference in a small region; on the other hand, two subordinate categories of close colors can only be well separated by texture. Furthermore, viewpoint, scale variation and occlusions among objects all contribute to the difficulties in this task [31]. Due to these challenges, similarity networks, which represent pair-wise affinity between images, can be very noisy and ineffective in retrieval of a sample from the correct species for any query.

We test our method on the Leeds butterfly fine-grained species image dataset [32]. Leeds Butterfly dataset contains 832 butterflies in 10 different classes with each class containing between 55 to 100 images [32]. We use two different common encoding methods (Fisher Vector (FV) and Vector of Linearly Aggregated Descriptors (VLAD) with dense SIFT; **Supplementary Note 1**) to generate two different vectorizations of each image. These two encoding methods describe the content of the images differently and therefore can contain different information. Each descriptor can generate a similarity network in which nodes represent images while edge weights indicate similarity between pairs of images. The inner product of these two similarity networks is used as the single input network to other denoising algorithms.

Visual inspection indicates that NE is able to greatly improve the overall similarity network for fine-grain identification (Figure A). While both encodings partially separate the species, before applying NE, all the images are tangled together without a clear clustering. On the other hand, the resulting similarity network after applying NE clearly shows 10 clusters corresponding to different butterfly species (Figure A). More specifically, given a certain query, the original input networks fail to capture the true affinities between the query butterfly and its most similar retrievals, while NE is able to correct the affinities and more reliably output the correct retrievals (Figure B).

To quantify the improvements due to NE in the task of species identification, we use identification accuracy, a standard metric which quantifies the average numbers of correct retrievals given any query of interests (**Supplementary Note 2**). A detailed comparison between NE and other alternatives by examining identification accuracy of the final network with respect to different number of top retrievals demonstrates NE's ability in improving the original noisy networks (Figure B). For example, when considering top 40 retrievals, NE can improve the raw network by 18.6% (more than 10% better than other alternatives). Further, NE generates the most significant improvement in performance (41% over the raw network and more than 25% over the second best alternative), when examining the top 80 retrieved images.

Current denoising methods suffer from high sensitivity to the hyper-parameters when constructing the input similarity networks, e.g., the variance used in Gaussian kernel (**Supplementary Note 1**). However, our model is more robust to the choice of hyper-parameters (**Supplementary Figure 3**). This robustness is due to the strict structure enforced by the preservation of symmetry and DSM structure during the diffusion process (see **Supplementary Note 3**).

# Discussion

We proposed Network Enhancement as a general method to denoise weighted undirected networks. NE implements a dynamic diffusion process that uses both local and global network structures to construct a denoised network from its noisy version. The core of our approach is a symmetric, positive semi-definite, doubly stochastic matrix, which is a theoretically justified replacement for the commonly used row-normalized transition matrix [33]. We showed that NE's diffusion model preserves the eigenvectors and increases the eigengap of this matrix for large eigenvalues. This finding provides insight into the mechanism of NE's diffusion and explains its ability to improve network quality. [15,16] In addition to increasing the eigengap, NE disproportionately trims small eigenvalues. This property can be contrasted with the principal component analysis (PCA) where the eigenspectrum is truncated at a particular threshold. Through extensive experimentation, we show that NE can flexibly fit into important network analytic pipelines in biology, and that its theoretical properties enable substantial improvements in performance of downstream network analyses.

We see many opportunities to improve upon the foundational concept of NE in future work. First, in some cases, a small subset of high confidence nodes may be available. For example, genomic regions in the Hi-C contact maps can be augmented using data obtained from 3C technology or a small number of species can be identified by a domain expert and used together with network data as input to a denoising methodology. Extending NE to take advantage of small amount of accurately labeled data might further extend our ability to denoise networks. Second, although we showed the utility of NE for denoising several types of weighted networks, there are other network types worth exploring, such as multimodal networks involving multiomic measurements of cancer patients. Finally, incorporating NE's diffusion process into other network analytic pipelines can potentially improve performance. For example, Mashup [17] learns vector representations for nodes based on a steady state of a traditional random walk with restart, and replacing Mashup's diffusion process with the rescaled steady state of NE might be a promising future direction.

## Methods

**Problem definition and doubly stochastic matrix property**

Let $G = (E, V, W)$ be a weighted network where $V$ denotes the set of nodes in the network (with $|V| = n$), $E$ represents the edges of $G$, and $W$ contains the weights on the edges. The goal of network enhancement is to generate a network $G^* = (E^*, V, W^*)$ that provides a better representation of the underlying module membership than the original network $G$. For the analysis below, we let $W$ represent a symmetric, non-negative matrix.

Diffusion-based models often rely on the row-normalized transition probability matrix $P = D^{-1}W$, where $D$ is a diagonal matrix whose entries are $D_{i,i} = \sum_{j=1}^{n} W_{i,j}$. However, transition probability matrix $P$ defined in this way is generally asymmetric and does not induce a directly usable node-node similarity metric. Additionally, most diffusion-based models lack spectral analysis of the denoised model. To construct our diffusion process and provide a theoretical analysis of our model, we propose to use a symmetric, doubly stochastic matrix (DSM). Given a matrix $M \in \mathbb{R}^{n \times n}$, $M$ is said to be DSM if:

1. $M_{i,j} \geq 0 \quad i,j \in \{1, 2, \ldots, n\}$,
2. $\sum_i M_{i,j} = \sum_j M_{i,j} = 1$.

The second condition above is equivalent to $\mathbf{1} = (1, 1, \ldots, 1)^T$ and $\mathbf{1}^T$ being a right and left eigenvector of $M$ with eigenvalue $1$. In fact, $1$ is the greatest eigenvalue for all DSM matrices (see the remark following the definition of DSM in the Supplementary Notes). Overall, the DSM property imposes a strict constraint on the scale of the node similarities and provides a scale-free matrix that is well-suited for subsequent analyses.

**Network Enhancement (NE)**

Given a matrix of edge weights $W$ representing the pairwise weights between all the nodes, we construct another localized network $\mathcal{T} \in \mathbb{R}^{n \times n}$ on the same set of nodes to capture local structures of the input network. Denote the set of $K$-nearest neighbors (KNN) of the $i$-th node (including the node $i$) as $\mathcal{N}_i$. We use these nearest neighbors to measure local affinity. Then the corresponding localized network $\mathcal{T}$ can be constructed from the original weighted network using the following two steps:

$$P_{i,j} \leftarrow \frac{W_{i,j}}{\sum_{k \in \mathcal{N}_i} W_{i,k}} \mathbb{I}_{\{j \in \mathcal{N}_i\}}, \qquad \mathcal{T}_{i,j} \leftarrow \sum_{k=1}^n \frac{P_{i,k} P_{j,k}}{\sum_{v=1}^n P_{v,k}}, \qquad (1)$$

where $\mathbb{I}_{\{\cdot\}}$ is the indicator function. We can verify that $\mathcal{T}$ is a symmetric DSM by directly checking the conditions of the definition (**Supplementary Note 3**). $\mathcal{T}$ encodes the local structures of the original network with the intuition that local neighbors (highly similar pairs of nodes) are more reliable than remote ones, and local structures can be propagated to non-local nodes through a diffusion process on the network. Motivated by the updates introduced in Zhou *et al.* [34], we define our diffusion process using $\mathcal{T}$ as follows:

$$W_{t+1} = \alpha \mathcal{T} \times W_t \times \mathcal{T} + (1-\alpha) \mathcal{T} \qquad (2)$$

where $\alpha$ is a regularization parameter and $t$ represents the iteration step. The value of $W_0$ can be initialized to be the input matrix $W$. Eqn. (2) shows that diffusion process in NE is defined by random walks of length three or less and a form of regularized information flow. There are three main reasons for restricting the influence of random walks to at most third-order neighbors in the network: (1) for most nodes third-order neighborhood spans the extent of almost the entire biological network, making neighborhoods of order beyond three not very informative of individual nodes [35,36], (2) currently there is little information about the extent of influence of a node (i.e., a biological entity, such as gene) on the activity (e.g., expression level) of its neighbor that is more than three hops away [37], and (3) recent studies have empirically demonstrated that network features extracted based on three-hop neighborhoods contain the most useful information for predictive modeling [38].

To further explore Eqn. (2) we can write the update rule for each entry:

$$(W_{t+1})_{i,j} = \alpha \sum_{k \in \mathcal{N}_i} \sum_{l \in \mathcal{N}_j} \mathcal{T}_{i,k}(W_t)_{k,l}\mathcal{T}_{l,j} + (1-\alpha)\mathcal{T}_{i,j}. \tag{3}$$

It can be seen from Eqn. (3) that the updated network comes from similarity/interaction flow only through the neighbors of each data point. The parameter $\alpha$ adds strengths to self-similarities, i.e., a node is always most similar to itself. One key property that differentiates our method from typical diffusion methods is that in the proposed diffusion process defined in Eqn. (2), for each iteration $t$, $W_t$ remains a symmetric DSM. Furthermore, $W_t$ converges to a non-trivial equilibrium network which is a symmetric DSM as well (**Supplementary Note 3**). Therefore, Network Enhancement constructs an undirected network that preserves the symmetry and DSM property of the original network. Through extensive experimentation we show that NE improves the similarity between related nodes and the performance of downstream methods such as community detection algorithms.

The main theoretical insight into the operation of NE is that the proposed diffusion process does not change eigenvectors of the initial DSM while mapping eigenvalues via a non-linear function (**Supplementary Note 3**). Let eigen-pair $(\lambda_0, \mathbf{v}_0)$ denote the eigen-pair of the initial symmetric DSM, $\mathcal{T}_0$. Then, the diffusion process defined in Eqn. (2) does not change the eigenvectors, and the final converged graph has eigen-pair $(f_\alpha(\lambda_0), \mathbf{v}_0)$, where $f_\alpha(x) = \frac{(1-\alpha)x}{1-\alpha x^2}$. This property shows that, the diffusion process using a symmetric, doubly stochastic matrix is a non-linear operator on the spectrum of the eigenvalues of the original network. This results has a number of consequences. Practically, it provides us with a closed-form expression for the converged network. Theoretically, it hints at how this diffusion process effects the eigen-spectrum and improves the network for subsequent analyses. 1) If the original eigenvalue is either $0$ or $1$, the diffusion process preserves this eigenvalue. This implies that, like other diffusion processes, NE does not connect disconnected components. 2) NE increases the gap between large eigenvalues of the original network and reduces the gap between small eigenvalues of this matrix. Larger eigengap is associated with better network community detection and higher-order network analysis [12,15,16]. 3) The diffusion process always decreases the eigenvalues, which follows from: $(1-\alpha)\lambda_0/(1-\alpha\lambda_0^2) \leq \lambda_0$, where smaller eigenvalues get reduced at a higher rate. This observation can be interpreted in relation to principal component analysis (PCA) where the eigenspectrum below a user determined threshold value is ignored. PCA has many attractive theoretical properties, especially for dimensionality reduction. In fact, Mashup [17], a feature learning method whose output is also a denoised version of the original network, can be fit by computing the PCA decomposition on the stationary state of the network. Mashup aims to learn a low-dimensional representation of nodes in the network which makes PCA a natural choice. However, a smoothed-out version of the PCA is more attractive for network denoising because denoising is typically used as a preprocessing step for downstream prediction tasks, and thus robustness to selection of a threshold value for the eigenspectrum is desirable.

These findings shed light on why the proposed algorithm (NE) enhances the robustness of the diffused network compared to the input network (**Supplementary Note 3**). In some contexts, we may need the output to remain a network of the same scale as the input network. This requirement can be satisfied by first recording the degree matrix of the input network and eventually mapping the denoised output of the algorithm back to the original scale by a symmetric matrix multiplication. We summarize our Network Enhancement algorithm along with this optional degree-mapping step in **Supplementary Note 3**.

**Code and Data Availability**

All relevant data are public and available from the authors of the original publications. The project website can be found at: http://snap.stanford.edu/ne. The website contains preprocessed data used in the paper together with raw and enhanced networks. Source code of the NE method is available for download from the project website.

**Acknowledgements.** M.Z. and J.L. were supported by NSF, NIH BD2K, DARPA SIMPLEX, Stanford Data Science Initiative, and Chan Zuckerberg Biohub. J.Z. was supported by the Stanford Graduate Fellowship, NSF DMS 1712800 Grant and the Stanford Discovery Innovation Fund. A.P. and C.D.B. were supported by National Institutes of Health/National Human Genome Research Institute T32 HG-000044, Chan Zuckerberg Initiative and Grant Number U01FD004979 from the FDA, which supports the UCSF-Stanford Center of Excellence in Regulatory Sciences and Innovation.

**Author information.** The authors declare no conflict of interest. Correspondence should be addressed to J.L. (jure@cs.stanford.edu) and to S.B. (serafim@cs.stanford.edu).

**Author contribution.** B.W., A.P., M.Z., and J.Z. designed and performed research, contributed new analytic tools, analyzed data, and wrote the manuscript. C.D.B., S.B., and J.L. designed and supervised the research and contributed to the manuscript.

**Additional information.** **Supplementary Information** contains a detailed description of datasets (**Supplementary Note 1**), additional experiments (**Supplementary Figures** and **Supplementary Note 2**), mathematical derivation, and theoretical analysis of NE (**Supplementary Note 3**).

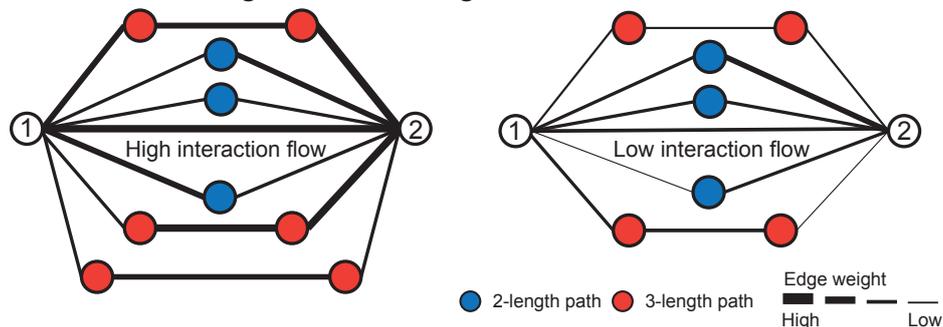
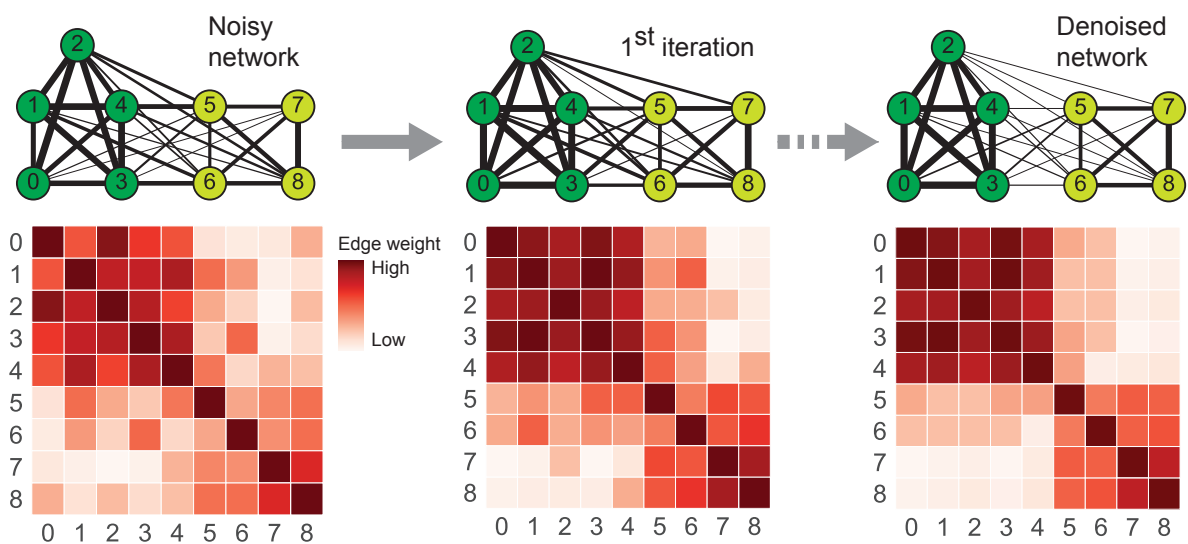

Figure 1: **Overview of Network Enhancement (NE).** (**A**) NE employs higher-order network structures to enhance a given weighted biological network. The diffusion process in NE revises edge weights in the network based on interaction flow between any two nodes. Specifically, for any two nodes, NE updates the weight of their edge by considering all paths of length three or less connecting those nodes. (**B**) The iterative process of NE. NE takes as input a weighted network and the associated adjacency matrix (visualized as a heat map). It then iteratively updates the network using the NE diffusion process, which is guaranteed to converge. The diffusion defined by NE improves the input network by strengthening edges that are either close to other strong edges in the network according to NE's diffusion distance or are supported by many weak edges. On the other hand, NE weakens edges that are not supported by many strong edges. Upon convergence, the enhanced network is a symmetric, doubly stochastic matrix (DSM) (**Supplementary Note 3**). This makes the enhanced network well-suited for downstream computational analysis. Furthermore, enforcement of the DSM structure leads to a more sparse networks with lower noise levels.

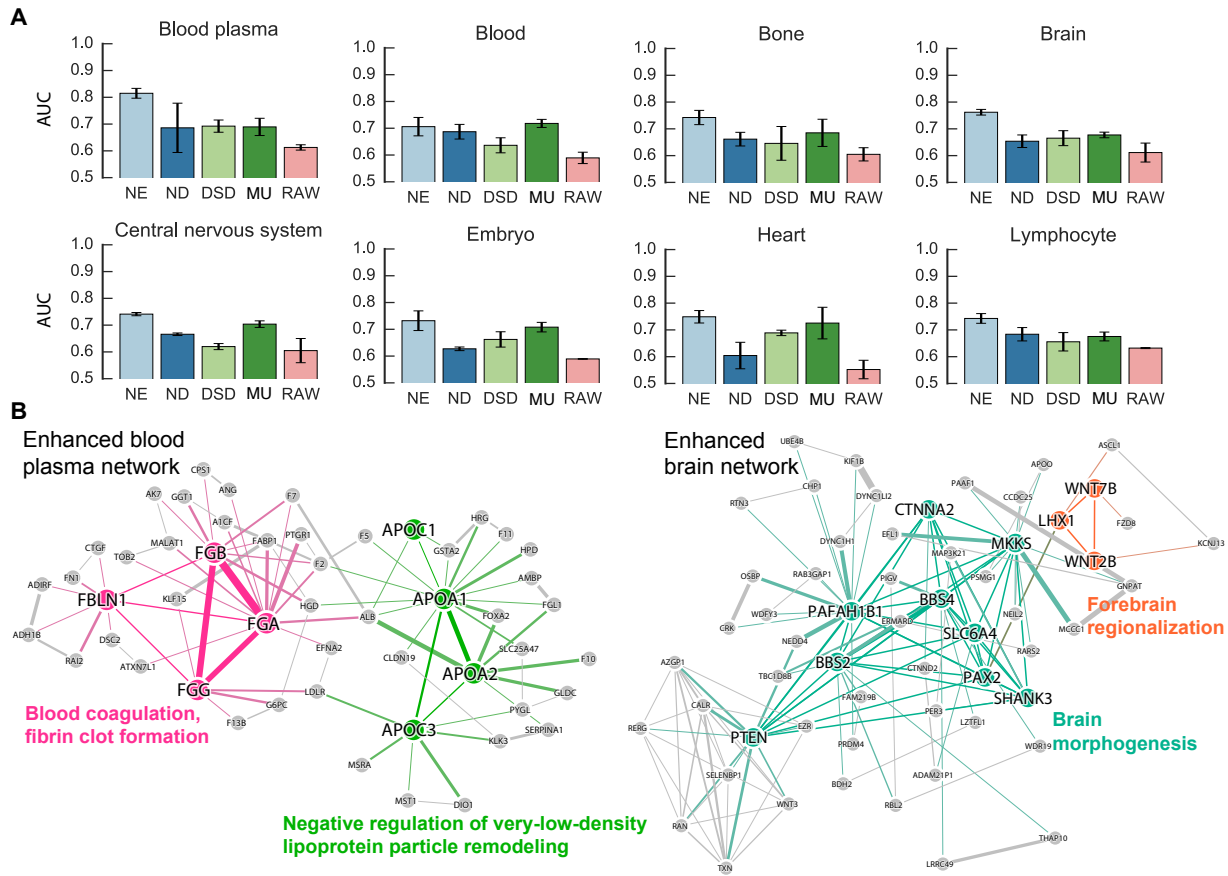

Figure 2: **Gene function prediction using tissue-specific gene interaction networks.** (**A**) We assessed the utility of original networks (RAW) and networks denoised using MU, ND, DSD and NE for tissue-specific gene function prediction. Each bar indicates the performance of a network-based approach that was applied to a raw or denoised gene interaction network in a particular tissue and then used to predict gene functions in that tissue. Prediction performance is measured using the area under receiver operating characteristic curve (AUROC), where a high AUROC value indicates the approach learned from the network to rank an actual association between a gene and a tissue-specific function higher than a random gene, tissue-specific function pair. Error bars indicate performance variation across tissue-specific gene functions. Results are shown for eight human tissues, additional fourteen tissues are considered in **Supplementary Figures 1 and 2**. (**B**) For blood plasma and brain tissues, we show gene interaction subnetworks centered on two blood plasma gene functions and two brain gene functions with the highest edge density in NE-denoised data. Edge density for each gene function (with $n$ associated genes) was calculated as the sum of edge weights in the NE-denoised network divided by the total number of possible edges between genes associated with that function ($n \times (n-1)/2$). The most interconnected gene functions in each tissue (shown in color, names of associated genes are emphasized), are also relevant to that tissue.

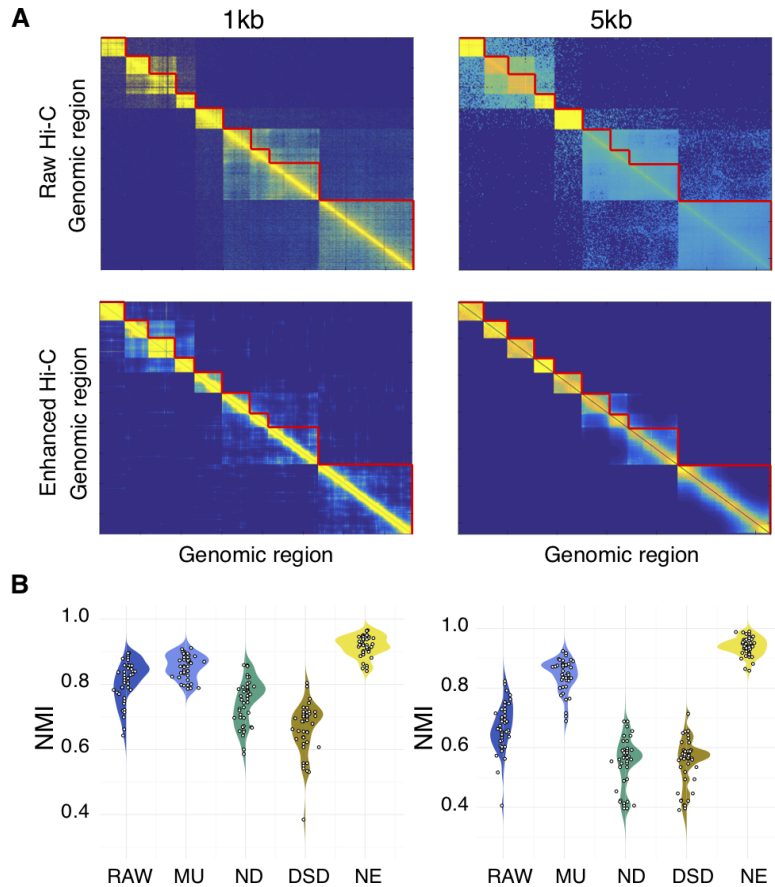

Figure 3: **Domain identification in Hi-C genomic interaction networks.** (**A**) Heatmap of Hi-C contact matrix for a portion of chromosome 16. For 1kb resolution data denoised with NE and clustered using Louvain community detection (**Supplementary Note 1**) chromosome 16 (visualized) has the median normalized mutual information (NMI), therefore, it was chosen as a fair representation of the overall performance. The top two heatmaps show the contact matrices for original (raw) data and the bottom heatmaps represent the contact matrices for data after application of NE. The images on the left correspond to data with 1kb resolution (*i.e.*, the bin-size is a 1kb region) and the right images correspond to the same section at 5kb resolution. The red lines indicate the borders for each domain as detailed in **Supplementary Note 1**. In each case, the network is consisted of genomic windows of length 1kb (left) or 5kb (right) as nodes, and normalized number of reads mapped to each region as the edge weights. The data was truncated for visualization purposes. (**B**) NMI for clusters detected. For each algorithm, the left side of the violin plot corresponds to Louvain community detection algorithm and the right side corresponds to MSCD algorithm. Each dot indicates the performance on a single autosome (The distance of the dots from the central vertical axis is dictated by a random jitter for visualization purposes). While for raw data and data preprocessed with DSD and ND the overall NMI decreases as resolution decreases, for NE and MU the performance remains high. MU maintains good overall performance with lower resolution, however, the spread of the NMI increases indicates that the consistency of performance has decreased compared to NE where the spread remains the same.

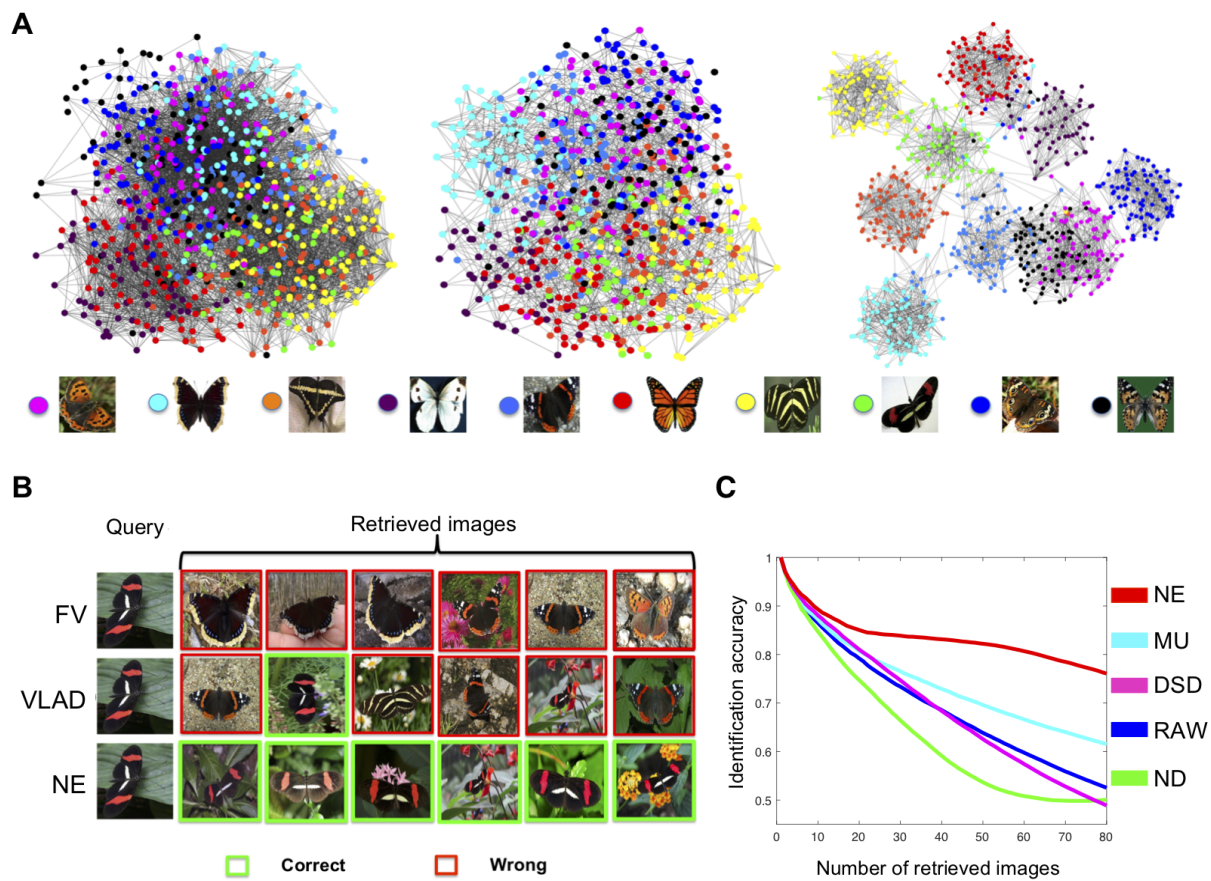

Figure 4: **Network-based butterfly species identification.** (Best seen in color.) Example of combining different metrics to improve the retrieval performance. (**A**) Visualization of encoded images as a network. From left to right: Fisher Vector, VLAD (**Supplementary Note 1**), and the denoised similarity network by our method (NE). The legend shows an example of each species included in the network. (**B**) Retrieval by each encoding method. Given a query butterfly, original descriptors fail to retrieve the correct species while the network denoised by NE is able to recover the correct similarities between the query and its neighbors within the same class. (**C**) Species identification accuracy when varying the number of retrieved images. A detailed comparison with other methods. Each curve shows the identification accuracy (**Supplementary Note 2**) as a function of number of retrievals for one method.

# Network Enhancement: a general method to denoise weighted biological networks

# Supplementary information

Bo Wang,[1,*] Armin Pourshafeie,[2,*] Marinka Zitnik,[1,*] Junjie Zhu,[3]
Carlos D. Bustamante,[4,5] Serafim Batzoglou,[1,‡,#] and Jure Leskovec[1,5,#]

[1] Department of Computer Science, Stanford University, Stanford, CA, USA
[2] Department of Physics, Stanford University, Stanford, CA, USA
[3] Department of Electrical Engineering, Stanford University, Stanford, CA, USA
[4] Department of Biomedical Data Science, Stanford University, Stanford, CA, USA
[5] Chan Zuckerberg Biohub, San Francisco, CA, USA
[‡] Currently at Illumina Inc.

[*] These authors contributed equally.

[#] To whom correspondence should be addressed. E-mail: serafim@cs.stanford.edu, jure@cs.stanford.edu

# Supplementary Figures

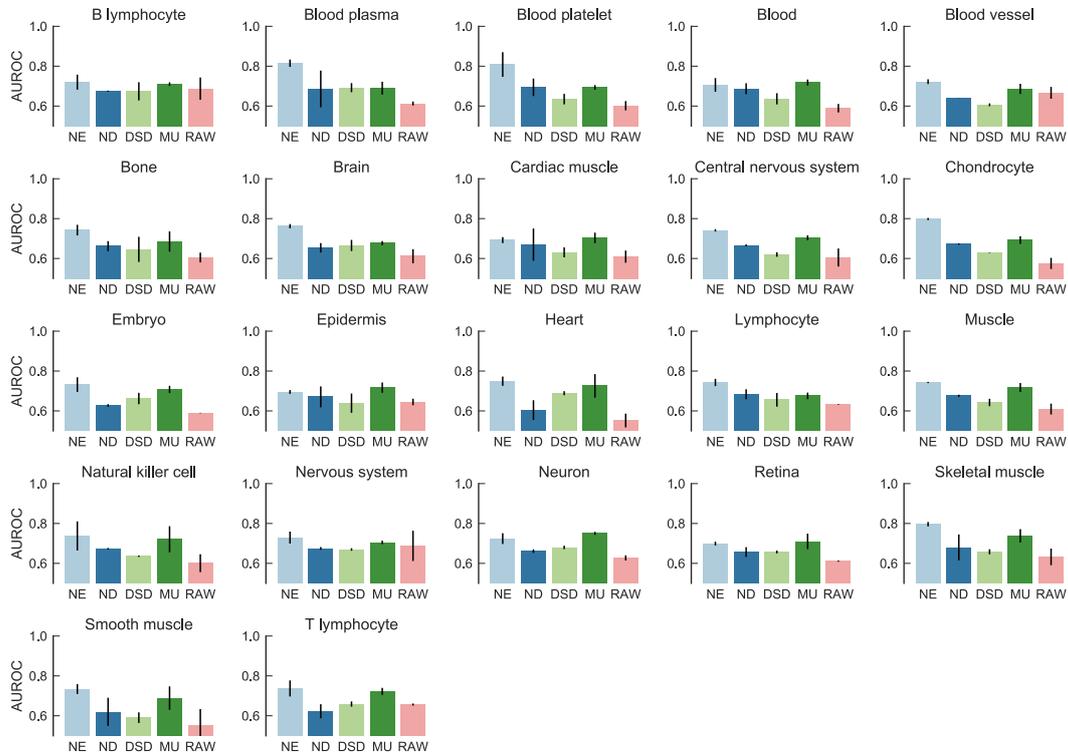

Figure 1: **Gene function prediction for genome-wide gene interaction networks in human tissues using a leave-one-out cross-validation setting.** We considered tissue-specific gene interaction networks [1] (RAW) and their denoised versions, which we obtained by applying MU, ND, DSD or NE to the original (RAW) networks. We then used the networks to predict gene functions specific to each tissue as defined in Greene *et al.* [1]. Each bar indicates the performance of a random-walk based approach that was applied to a raw or a denoised network in order to predict gene functions taking place in the tissue described by the network. Prediction performance is measured using AUROC, where a high AUROC value indicates the approach successfully learned to rank an actual gene-function association higher than a random gene-function pair. Error bars indicate performance variation across all gene functions in a given tissue. Results are shown for all 22 human tissues considered in this study. The average AUROC values achieved by the methods across 22 tissues are: NE: 0.742, ND: 0.662, DSD: 0.649, MU: 0.706, and RAW: 0.616.

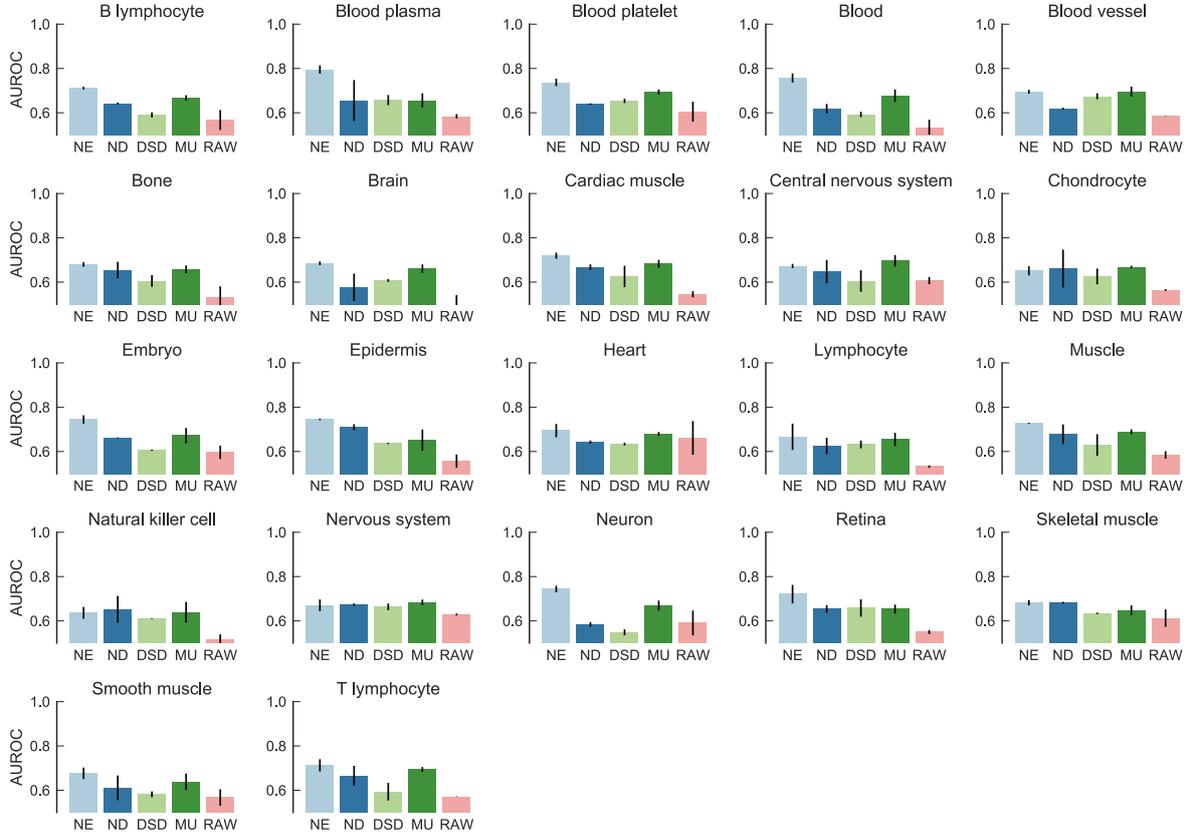

Figure 2: **Gene function prediction for genome-wide gene interaction networks in human tissues using a 5-fold cross-validation setting.** We considered tissue-specific gene interaction networks [1] (RAW) and their denoised versions, which we obtained by applying MU, ND, DSD or NE to the original (RAW) networks. We then used the networks to predict gene functions specific to each tissue as defined in Greene *et al.* [1]. Each bar indicates the performance of a random-walk based approach that was applied to a raw or a denoised network in order to predict gene functions taking place in the tissue described by the network. Prediction performance is measured using AUROC, where a high AUROC value indicates the approach successfully learned to rank an actual gene-function association higher than a random gene-function pair. Error bars indicate performance variation across all gene functions in a given tissue. Results are shown for all 22 human tissues considered in this study. The average AUROC values achieved by the methods across 22 tissues are: NE: 0.706, ND: 0.646, DSD: 0.621, MU: 0.669, and RAW: 0.572.

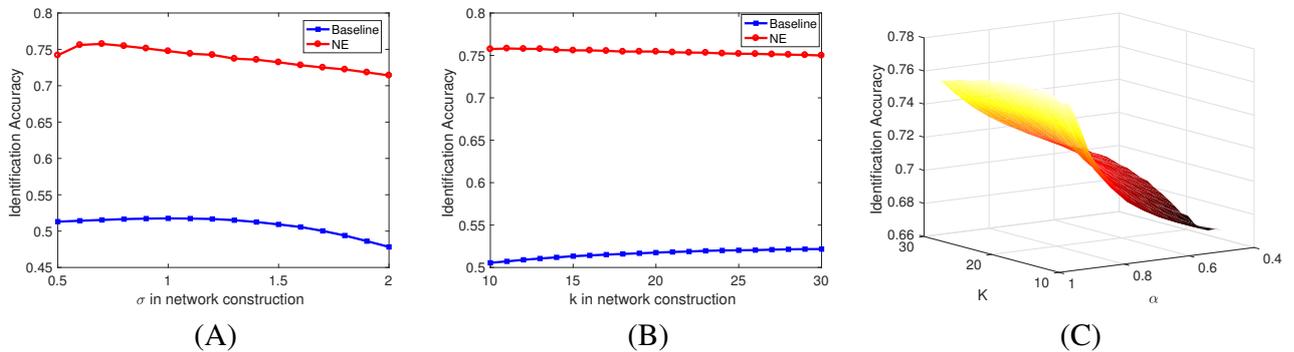

Figure 3: **Results of NE on fine-grained species identification.** (**A**) shows the sensitivity to the hyper-parameter, $\sigma$, when constructing the similarity network on butterfly dataset. (**B**) shows the sensitivity to the hyper-parameter $k$ for this network. (**C**) is a mesh plot of different values of $K$ and $\alpha$ for our network enhancement on the butterfly dataset.

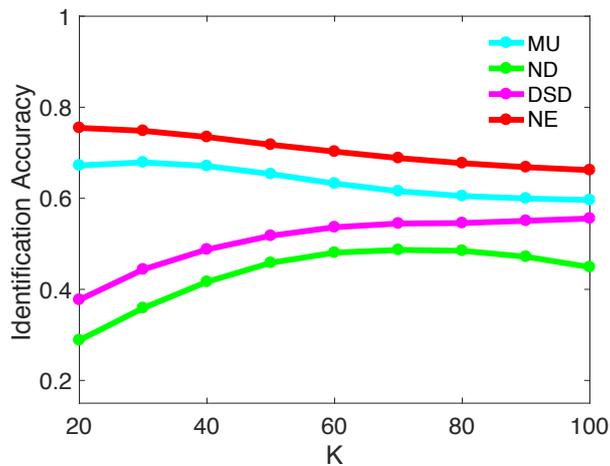

Figure 4: **Species Identification Accuracy with respect to different number of $K$ in K-NN pruning as a pre-processing step.** We perform the same KNN pruning for all methods and report the corresponding identification accuracy. It is observed that, NE outperforms the alternative methods for various choices of $K$. Furthermore, both NE and MU perform better for smaller $K$, the performance of DSD improves as $K$ is increased and ND performs best at an intermediate value in the range investigated.

# Supplementary Note 1: Further information on datasets

**Tissue-Specific Gene Interaction Networks**

Tissue-specific gene interaction networks were retrieved from the GIANT (Genome-scale Integrated Analysis of gene Networks in Tissues) database [1]: http://giant.princeton.edu. Networks were filtered to only include edges with evidence supporting a tissue-specific functional interaction (*i.e.,* network type "top edges" in the GIANT database). Each network was used as input to a network denoising algorithm to clean the network edges. The resulting denoised network was then used as input to a random-walk based algorithm to predict gene functions.

Gene functions were defined by the Gene Ontology (GO) terms [2]. Gene-function associations were specified by the GO annotations [2] and retrieved from ftp://ftp.ncbi.nlm.nih.gov/gene/DATA/gene2go.gz in August 2016. We only used high confidence annotations associated with the experimental evidence codes: EXP, IDA, IMP, IGI, IEP, ISS, ISA, ISM or ISO, and further removed all annotations with a non-empty "qualifier" column [3]. The original GO files only contained the most specific annotations explicitly. We therefore added all implicit more general annotations by up-propagating the given annotations along the full GO tree.

We obtained the mapping of GO terms to tissues, that is, associations between tissues and tissue-specific functions, from Greene *et al.* [1]. Greene *et al.* used text matching followed by manual curation to map GO terms to tissues. GO terms were filtered to only include those with at least 20 associated genes. As a result, there were 22 tissues with each having at least one tissue-specific gene function. In total, there were 309 tissue-specific gene functions across all 22 tissues. Tissues with the largest number of functions were: natural killer cell (49 GO terms), lymphocyte (43 GO terms) and muscle (34 GO terms).

To predict gene functions we used a random-walk based approach. Random walks were used before to transfer GO annotations within networks ([4–8] and many others) and were shown to be among the top-perfoming approaches for gene function prediction [5,6]. We defined a random walk starting from nodes that were known to be associated with a query gene function and were included in the training set. At each time step, the walk had a probability $r$ of returning to the initial nodes. We set $r = 0.75$, as was done by Köhler *et al.* [4]. Once the random walk process converged (L2-distance between probability vectors in consecutive time steps $< 10^{-6}$), predictions were made for all nodes in the test set based on their visitation probability. Predictions were evaluated against known gene-function associations using a leave-one-out cross-validation strategy.

**Hi-C Interaction Networks**

For each autosome, the provided contact matrix (counts per bin) from Rao et al. [9] was normalized using SQRTVC as defined in [9]. In their work Rao et al. also introduced the Arrowhead algorithm as a way of detecting clusters within a Hi-C adjacency matrix. The Arrowhead algorithm produces clusters that may overlap. Since the true clusters are unknown, to generate a confident set of labels, for each chromosome, we sub-sampled the first 15 non-overlapping clusters that contain no sub-clusters as determined by the arrowhead algorithm [9]. We chose non-overlapping clusters as both of the community detection algorithm we use for post processing are limited to detecting non-overlapping communities [10,11]. This sub-sampled adjacency matrix constitutes the contact matrix for our new Hi-C interaction network. For visualization purposes, we only show the first 9 communities. We have chosen chromosome 16 as our visualization example. This example was chosen to have a performance just below the median as measured by NMI of Louvian clustering for 1kb resolution Hi-C.

**Fine-Grained Image Datasets and Similarity Networks**

First, we test our method on a dataset with 10 different classes of butterflies, each of which containing 55 to 100 images totaling to 832 butterflies [12]. We use two different encoding methods (Fisher Vector (FV) [13,14] and Vector of Linearly Aggregated Descriptors (VLAD) [15] with dense SIFT [16]) to generate two different descriptors for these images. These two encoding methods describe the statistics of the codebooks differently and therefore we use our method to combine them.

Given a feature set that describes a collection of images, denoted as $X = \{x_1, x_2, \ldots, x_n\}$, we want to construct a similarity graph $\mathcal{N} \in R^{n \times n}$ in which $\mathcal{W}(i,j)$ indicates the kernel value between the $i$-th and $j$-th object. The most widely used method assumes a Gaussian distribution across pairwise similarities:

$$\mathcal{W}(i,j) = \exp\left(-\frac{\|x_i - x_j\|^2}{2\sigma^2}\right).$$

Here, $\sigma$ is a hyper-parameter that needs careful manual tuning. To overcome the sensitivity to $\sigma$, a more advanced method of constructing similarity kernels is proposed in [17] where the variance is estimated using the local scales of the distances as follows. Assume $k$ is the number of neighbors. For each cell, e.g, $x_i$, the associated local variance is estimated as:

$$\epsilon_i = \frac{\sum_{j \in \mathcal{KNN}(i)} \|x_i - x_j\|}{k},$$

where $\mathcal{KNN}(i)$ denotes all the top $k$ neighbors of the $i$-th cell. Thus the new kernel is defined as:

$$\mathcal{W}_k^\sigma(i,j) = \exp\left(-\frac{\|x_i - x_j\|^2}{\sigma^2(\epsilon_i + \epsilon_j)^2}\right).$$

We set $k = 20$ and $\sigma = 0.5$ as default values.

## Supplementary Note 2: Definition of Evaluation Metrics

**Normalized Mutual Information**

Throughout the paper, we used Normalized Mutual Information (NMI) [18] to evaluate the consistency between the obtained clustering and the true labels of the cells. Given two clustering results $U$ and $V$ on a set of data points, NMI is defined as: $I(U,V)/\max\{H(U), H(V)\}$, where $I(U,V)$ is the mutual information between $U$ and $V$, and $H(U)$ represents the entropy of the clustering $U$.

Specifically, assuming that $U$ has $P$ clusters, and $V$ has $Q$ clusters, the mutual information is computed as follows:

$$I(U,V) = \sum_{p=1}^{P}\sum_{q=1}^{Q} \frac{|U_p \cap V_q|}{N} \log \frac{|U_p \cap V_q|}{|U_p|/N \times |V_q|/N},$$

where $N$ is the number of points and $|U_p|$ denotes the cardinality of the $p$-th cluster in $U$. The entropy of each cluster assignment is calculated as follows:

$$H(U) = -\sum_{p=1}^{P} \frac{|U_p|}{N} \log \frac{|U_p|}{N},$$

$$H(V) = -\sum_{q=1}^{Q} \frac{|V_q|}{N} \log \frac{|V_q|}{N}.$$

Further details on NMI can be found in Vinh et al.[19]. NMI takes on values between $0$ and $1$ where a higher NMI indicates a higher concordance between the two sets, *i.e.,* a more consistent label assignment.

**Retrieval Accuracy**

We use retrieval accuracy for evaluation of fine-grained image retrieval. For a single query $q$, the accuracy on $k$ retrievals is defined as:

$$acc(q,k) = \frac{\text{\# of correct retrievals}}{min(k, N_q)},$$

where $N_q$ is the number of objects with the same label of $q$. Here, "correct retrievals" mean the retrieved images from the same class of $q$. We also report the mean accuracy $Acc$ over all the images in the dataset:

$$Acc = \frac{1}{n}\sum_{i=1}^{n} acc(q_i, N_{q_i}),$$

where $n$ is the number of images in the dataset.

# Supplementary Note 3: Theoretical analysis of Network Enhancement
## Doubly Stochastic Matrix

Here, we state the definition of a Doubly Stochastic Matrix (*DSM*):

**Definition 1.** Given a matrix $M \in \mathbb{R}^{n \times n}$, $M$ is a *Doubly Stochastic Matrix (DSM)* if it satisfies the following two conditions:
1. $M_{i,j} \geq 0 \quad i,j \in \{1, 2, \ldots, n\}$,
2. $\sum_i M_{i,j} = \sum_j M_{i,j} = 1$.

*Remark.* The largest eigenvalue of a DSM matrix is 1. It is easy to check that $\mathbf{1} = (1, 1, \ldots 1)^T$ is a right eigenvector with eigenvalue 1. Similarly, $\mathbf{1}^T$ is a left eigenvector with eigenvalue 1. For an irreducible DSM, Perron-Frobenius theorem implies that the $(\mathbf{1}, 1)$ pair is unique and 1 is the largest eigenvalue. When $M$ is reducible, its indices can be split to construct $k$ irreducible DSM's. Any eigenvector of $M$ needs to be an eigenvector of all of these matrices. Since the eigenvalue corresponding to each of these matrices cannot be greater than 1 we conclude that the largest eigenvalue of a reducible DSM is 1 corresponding to eigenvector $\mathbf{1}$ and potentially other eigenvectors.

Next, we show that the transition matrix is a DSM. First we re-state the construction of transition matrices:

$$P_{i,j} \leftarrow \frac{W_{i,j}}{\sum_{k \in \mathcal{N}_i} W_{i,k}} * I\{j \in \mathcal{N}_i\}, \qquad \mathcal{T}_{i,j} \leftarrow \sum_{k=1}^{n} \frac{P_{i,k} P_{j,k}}{\sum_{v=1}^{n} P_{v,k}}. \tag{1}$$

Where $I\{\cdot\}$ is an indicator function. By checking the conditions from the definition of DSM, we verify that $\mathcal{T}$ is a symmetric DSM.

Given a weighted graph $W \in \mathbb{R}^{n \times n}$, the transition probability matrix $P = D^{-1}W$, where $D$ is the diagonal matrix whose entries are the degree of the vertices, i.e., $D_{ii} = \sum_{j=1}^{n} W_{i,j}$. In other words, we have:

$$P_{i,j} = \frac{W_{i,j}}{\sum_{k=1}^{n} W_{i,k}}. \tag{2}$$

It is easy to verify that, $P\mathbf{1} = \mathbf{1}$, *i.e.*, the row sum of $P$ is always 1. Note $P$ is not symmetric. Now we construct the DSM matrix $\mathcal{T}$ as follows:

$$\mathcal{T}_{i,j} \leftarrow \sum_{k=1}^{n} \frac{P_{i,k} P_{j,k}}{\sum_{v=1}^{n} P_{v,k}}. \tag{3}$$

It is easy to see that, $\mathcal{T} \in \mathbb{R}^{n \times n}$ is symmetric:

$$\mathcal{T}_{i,j} = \sum_{k=1}^{n} \frac{P_{i,k} P_{j,k}}{\sum_{v=1}^{n} P_{v,k}} = \sum_{k=1}^{n} \frac{P_{j,k} P_{i,k}}{\sum_{v=1}^{n} P_{v,k}} = \mathcal{T}_{j,i}.$$

It remains to show that $\mathcal{T}$ is a DSM.

Since weights are assumed to be non-negative, $W_{i,j} \geq 0$. This implies that $P_{i,j}$ as defined in equation 2 is non-negative and therefore, $\mathcal{T}_{i,j} \geq 0$

Next we show that the second property of DSM holds by by first proving $\mathcal{T}\mathbf{1} = \mathbf{1}$:

$$(\mathcal{T}\mathbf{1})_i = \sum_{j=1}^n \sum_{k=1}^n \frac{P_{i,k} P_{j,k}}{\sum_{v=1}^n P_{v,k}} = \sum_{k=1}^n P_{i,k} \frac{\sum_{j=1}^n P_{j,k}}{\sum_{v=1}^n P_{v,k}} = \sum_{k=1}^n P_{i,k} = 1. \quad (4)$$

This implies that, each row sum of $\mathcal{T}$ is 1, so $\mathcal{T}\mathbf{1} = \mathbf{1}$. If we take transpose on both sides, we have $\mathbf{1}'\mathcal{T}' = \mathbf{1}'$, and since $T$ is symmetric (i.e., $\mathcal{T}' = \mathcal{T}$), then we obtain $\mathbf{1}'\mathcal{T} = \mathbf{1}'$. So, we conclude that the row sums and the column sums of $\mathcal{T}$ are always 1. This proves that $\mathcal{T}$ is a DSM. Put together, we have that $\mathcal{T}$ is symmetric doubly stochastic matrix.

Further, we can see that $\mathcal{T}$ will be positive semi-definite. To show this, for any vector $\mathbf{z}$, we need to prove $\mathbf{z}'\mathcal{T}\mathbf{z} \geq 0$:

$$\begin{aligned}
\mathbf{z}'\mathcal{T}\mathbf{z} &= \sum_{i=1}^n \sum_{j=1}^n z_i z_j \mathcal{T}_{i,j} = \sum_{i=1}^n \sum_{j=1}^n z_i z_j \sum_{k=1}^n \frac{P_{i,k} P_{j,k}}{\sum_{v=1}^n P_{v,k}} \\
&= \sum_{k=1}^n \frac{\sum_{i=1}^n \sum_{j=1}^n z_i z_j P_{i,k} P_{j,k}}{\sum_{v=1}^n P_{v,k}} = \sum_{k=1}^n \frac{(\sum_{i=1}^n z_i P_{i,k})^2}{\sum_{v=1}^n P_{v,k}} \geq 0.
\end{aligned}$$

We thus confirmed that $\mathcal{T}$ is positive semi-definite.

Furthermore, we can easily verify that convex combinations of symmetric DSMs is still a symmetric DSM.

*Proof.* This proof follows immediately from the definition. Given $m$ DSMs, $A_i$, for $i = 1, 2, \ldots, m$, a convex combination is $\sum_i^m \beta_i A_i$, such that $\sum_i^m \beta_i = 1$ and $\beta_i \geq 0, i = 1, \ldots, m$. The symmetry of a convex combination of symmetric matrices is trivial. The first property of DSM follows since all values involved are non-negative and are added or multiplied. The second property is also easy to confirm using $(\sum_i \beta_i A_i)\mathbf{1} = \sum_i \beta_i (A_i \mathbf{1}) = \mathbf{1}$. Transposing this equation and using symmetry shows the results for the column sums. □

**Network Enhancement Preserves Properties of DSM**

Network enhancement diffusion process is given by:

$$W_{t+1} = \alpha \mathcal{T} \times W_t \times \mathcal{T} + (1-\alpha)\mathcal{T}, \quad (5)$$

where initialization is done by $W_{t=0} \leftarrow W$, with $\alpha$ a regularization parameter, and $t$ representing the iteration number.

**Theorem 1.** *In each iteration $t$ of network enhancement (NE) as defined by Eqn. (5), the following properties hold :*

1. *$W_t$ remains a symmetric DSM.*

2. *$W_t$ converges to a non-trivial equilibrium graph that is a symmetric DSM.*

3. *$W_t$ remains positive-semi definite if $W_{t=0}$ is positive semi-definite.*

*Proof.* To prove the first statement, we focus on checking the definitions. Given that $W_{t=0}$ and the local graph $\mathcal{T}$ are symmetric DSMs, we can proceed by induction on $t$. Assume $W_t$ is a symmetric DSM, we want to verify that $W_{t+1}$ is again symmetric and a DSM. We start by proving symmetry:

$$W'_{t+1} = \alpha(\mathcal{T}W_t\mathcal{T})' + (1-\alpha)\mathcal{T} = \alpha(\mathcal{T}'W'_t\mathcal{T}') + (1-\alpha)\mathcal{T} = \alpha(\mathcal{T}W_t\mathcal{T}) + (1-\alpha)\mathcal{T} = W_{t+1}.$$

Here, we use $W'_t = W_t$ and $\mathcal{T}' = \mathcal{T}$. Hence, $W_{t+1}$ is symmetric.

We proceed to show that $W_{t+1}$ remains doubly stochastic. It is obvious that each element of $W_{t+1}$ is non-negative. To show the rows and columns remain normalized, we note that:

$$W_{t+1}\mathbf{1} = \alpha\mathcal{T}W_t\mathcal{T}\mathbf{1} + (1-\alpha)\mathcal{T}\mathbf{1} = \alpha\mathcal{T}W_t\mathbf{1} + (1-\alpha)\mathcal{T}\mathbf{1} = \alpha\mathcal{T}\mathbf{1} + (1-\alpha)\mathcal{T}\mathbf{1} = \mathcal{T}\mathbf{1} = \mathbf{1}.$$

here we have used $\mathcal{T}\mathbf{1} = \mathbf{1}$ and $W_t\mathbf{1} = \mathbf{1}$, since they are both DSMs. This shows that $W_{t+1}$ is row normalized. We can appeal to symmetry to show that the matrix will also be column normalized which shows statement 1.

Next we show that it is possible to find a closed form solution for the final, converged network. We start by first providing an expression for the network at iteration $t$. Then we find the network in the limit of large number of iterations.

Define $W_0 = W_{t=0}$. For iteration $t$, the following holds true:

$$W_t = \alpha^t \mathcal{T}^t W_0 \mathcal{T}^t + (1-\alpha)\mathcal{T}\sum_{k=0}^{t-1}(\alpha\mathcal{T}^2)^k. \tag{6}$$

which can be shown by induction. For $t = 1$, $W_{t=1} = \alpha\mathcal{T}W_0\mathcal{T} + (1-\alpha)\mathcal{T}$, and clearly satisfies Eqn. (6). Assume Eqn. (6) is true for iteration $t$. Then:

$$\begin{aligned} W_{t+1} &= \alpha\mathcal{T}W_t\mathcal{T} + (1-\alpha)\mathcal{T} \\ &= \alpha\mathcal{T}(\alpha^t\mathcal{T}^tW_0\mathcal{T}^t + (1-\alpha)\mathcal{T}\sum_{k=0}^{t-1}(\alpha\mathcal{T}^2)^k)\mathcal{T} + (1-\alpha)\mathcal{T} \\ &= \alpha^{t+1}\mathcal{T}^{t+1}W_0\mathcal{T}^{t+1} + (1-\alpha)\mathcal{T}\sum_{k=0}^{t}(\alpha\mathcal{T}^2)^k. \end{aligned}$$

which satisfies Eqn. (6). Let $t \to \infty$, then:

$$W_{t\to\infty} = (1-\alpha)\mathcal{T}(\mathcal{I} - \alpha\mathcal{T}^2)^{-1}.$$

This proves that the network enhancement process converges to a non-trivial equilibrium graph $W_{t\to\infty} = (1-\alpha)\mathcal{T}(\mathcal{I}-\alpha\mathcal{T}^2)^{-1}$. Note that this result is the limit of symmetric DSM matrices. The set of symmetric $n \times n$ doubly stochastic matrices can be described by $\{M : M - M^T = 0, M_{i,j} \geq 0, \sum_i M_{i,j} = 1, \sum_j M_{i,j} = 1\}$. Since these conditions are inverse images of closed sets ($\{0\}, [0, \infty), \{1\}, \{1\}$ respectively) under continuous maps, the set of symmetric DSMs is closed and contains the limit point corresponding to the converged diffusion network in NE.

Lastly, we argue that if $W_0$ is positive semi-definite, then the NE diffusion process preserves this property at every iteration. By induction, let $W_t$ be positive semi-definite then for any vector $\mathbf{z} \in \mathbb{R}^n$:

$$\mathbf{z}'W_{t+1}\mathbf{z} = \alpha\mathbf{z}'\mathcal{T}W_t\mathcal{T}\mathbf{z} + (1-\alpha)\mathbf{z}'\mathcal{T}\mathbf{z} = \alpha(\mathcal{T}\mathbf{z})'W_t(\mathcal{T}\mathbf{z}) + (1-\alpha)\mathbf{z}'\mathcal{T}\mathbf{z} \geq 0.$$

Finally, we argue that since the set of positive semi-definite matrices can be represented by $\{M : f(M) \geq 0\}$ where $f(M) = \min_{\|x\|=1}\langle x, Mx \rangle$ is a continuous function, the set of positive semi-definite matrices is closed (and thus contains it's limit points) as it is the inverse image of $[0, \infty)$ under $f$. $\square$

This theorem demonstrates that the diffusion process in NE preserves some important properties of the original network. Importantly, at every stage of the diffusion process, the results corresponds to an undirected network with the same normalization as the initial network.

**Spectral Analysis of Network Enhancement**

Now we present our main novel finding that the proposed network enhancement process does not change eigenvectors of the initial symmetric DSM while mapping eigenvalues via a non-linear function.

**Theorem 2.** *Let $(\lambda_0, \mathbf{v}_0)$ denote the eigen-pair of a symmetric DSM $\mathcal{T}_0$. Then the network enhancement process defined in Eqn. (5) does not change the eigenvectors and the final converged graph has an eigen-pair $(f_\alpha(\lambda_0), \mathbf{v}_0)$, where $f_\alpha(x) = \frac{(1-\alpha)x}{1-\alpha x^2}$.*

*Proof.* Let $\mathcal{T}_0$ denote the initial symmetric DSM and $\mathcal{T}_\infty$ denote the final symmetric DSM. From the proof above, it is easy to see that the final network $\mathcal{T}_\infty$ is given by $\mathcal{T}_\infty = (1-\alpha)\mathcal{T}_0(\mathcal{I}-\alpha\mathcal{T}_0^2)^{-1}$. Since $\mathcal{T}_0$ is a symmetric DSM, then we have $\mathcal{T}_0 = U\Sigma U^{-1}$ where $U$ is the set of eigenvectors and $\Sigma$ is a diagonal matrix whose entries are eigenvalues of $\mathcal{T}_0$, i.e., $\Sigma_{i,i} = \lambda_i$. Clearly,

$$\begin{aligned}
\mathcal{T}_\infty &= (1-\alpha)\mathcal{T}_0(\mathcal{I} - \alpha\mathcal{T}_0^2)^{-1} \\
&= (1-\alpha)U\Sigma U^{-1}(\mathcal{I} - \alpha U\Sigma U^{-1}U\Sigma U^{-1})^{-1}. \\
&= (1-\alpha)U\Sigma U^{-1}(UU^{-1} - \alpha U\Sigma U^{-1}U\Sigma U^{-1})^{-1}. \\
&= (1-\alpha)U(\Sigma(\mathcal{I} - \alpha\Sigma^2)^{-1})U^{-1}.
\end{aligned}$$

Hence, we obtain the eigen-decompostion of $\mathcal{T}_\infty$. That is, the eigenvectors are still $U$ but the eigenvalues becomes $\Sigma'_{i,i} = (1-\alpha)\lambda_i(1-\alpha\lambda_i^2)^{-1}$. This completes the proof of the theorem. $\square$

This theorem shows that, the defined network enhancement process using a DSM is a nonlinear operator on the eigenvalue-spectrum of the network. This theorem not only provides us with a closed-form expression for obtaining the final network at convergence but also sheds light on how network enhancement process improves the graph. First, if the original eigenvalues are either $0$ or $1$, the network enhancement process preserves these eigenvalues. Second, network enhancement process always decreases the eigenvalues since $\frac{(1-\alpha)\lambda_0}{1-\alpha\lambda_0^2} \leq \lambda_0$. More importantly, NE increases the eigengaps between large eigenvalues (Lemma 1) and thereby enhances the robustness of the obtained graph (Theorem 3) and influences clustering. Third, while all eigenvalues are reduced, the non-linear function $f_\alpha$ reduces small eigenvalues more aggressively than large eigenvalues. In this sense, NE acts similar to a smoothed out version of PCA but does not completely diminish any singular value.

Consider the initial graph $\mathcal{T}_0 \in \mathbb{R}^{n \times n}$ and the obtained graph $\mathcal{T}_\infty \in \mathbb{R}^{n \times n}$ after the network enhancement process. Then,

**Lemma 1.** *Let, $c(\alpha) = \sqrt{-\frac{\sqrt{\alpha^2-10\alpha+9}+\alpha-3}{2\alpha}}$, for all eigenvectors with eigengap contained in $[1, c(\alpha)]$ (i.e. $\lambda_{i+1} \geq c(\alpha)$) the eigengap is larger in $\mathcal{T}_\infty$ than in $\mathcal{T}_0$.*

*Proof.* First we note that by Theorem 2, $\mathcal{T}_\infty, \mathcal{T}_0$ share the same eigenvectors. Let $k$ be the last eigenvector with $\lambda_{k+1} \geq c(\alpha)$. The lemma reduces to showing:

$$\|\lambda_j - \lambda_{j+1}\| \leq \|\lambda_j^{(\infty)} - \lambda_{j+1}^{(\infty)}\|, \text{ with } j \leq k$$

where $\lambda_j^{(\infty)}$ is the $j$-th eigenvalues of the final graph. By Theorem 2, we have $\lambda_j^{(\infty)} = \frac{(1-\alpha)\lambda_j}{1-\alpha\lambda_j^2}$, therefore, the preceding equations becomes:

$$\lambda_j - \frac{(1-\alpha)\lambda_j}{1-\alpha\lambda_j^2} \leq \lambda_{j+1} - \frac{(1-\alpha)\lambda_{j+1}}{1-\alpha\lambda_{j+1}^2}.$$

Since $\lambda_j \geq \lambda_{j+1}$, the claim holds where $g_\alpha(x) = x - \frac{(1-\alpha)x}{1-\alpha x^2}$ is a decreasing function. Differentiating $g_\alpha(x)$, gives the following condition:

$$\frac{\partial g_\alpha(x)}{\partial x} = 1 - (1-\alpha)\frac{(1+\alpha x^2)}{(1-\alpha x^2)^2} \leq 0.$$

Since $0 < \alpha < 1$, this condition implies that: $x^4 \alpha + x^2(\alpha - 3) + 1 \geq 0$, or that:

$$|x| \geq \sqrt{-\frac{\sqrt{\alpha^2 - 10\alpha + 9} + \alpha - 3}{2\alpha}} = c(\alpha).$$

□

One implication of this lemma is an increased robustness. For $H \in \mathbb{R}^{n \times n}$, a symmetric perturbation, define $\overline{M} := M + H$ as the perturbed version of $M$. Further, denote the eigenspace spanned by the largest $k$ eigenvectors of $M$ by $V_{M,k}$. Then, let $\mathbf{dist}(V_M, V_{\overline{M}})$ indicates the distance between projected eigenspaces of $M$ and $\overline{M}$ (see detailed definition in the review by Von Luxburg [20]).

**Theorem 3.** *(Perturbation Analysis)* $\mathcal{T}_\infty$ *has a better resistance to noise than* $\mathcal{T}_0$ *in the following sense:*

$$\sup_{\substack{\|H\|=h \\ \mathcal{T}_0}} \{\mathbf{dist}(V_{\mathcal{T}_0,k}, V_{\overline{\mathcal{T}_0},k})\} \geq \sup_{\substack{\|H\|=h \\ \mathcal{T}_0}} \{\mathbf{dist}(V_{\mathcal{T}_\infty,k}, V_{\overline{\mathcal{T}_\infty},k})\},$$

*for all $k$ with $\lambda_{k+1} \geq c(\alpha)$ in $\mathcal{T}_0$.*

To prove this theorem, the key observation lies in the fact that for large eigenvalues, the eigengap of $\mathcal{T}_0$ is always smaller than the corresponding eigengap of $\mathcal{T}_\infty$.

*Proof.* First, we directly use a modified version of Davis-Kahan theorem (Theorem 2 from [21]). The text of theorem 2 from [21] is reproduced below for completeness:

Let $\Sigma, \hat{\Sigma} \in \mathbb{R}^{p \times p}$ be symmetric, with eigenvalues $\lambda_1 \geq \cdots \geq \lambda_p$ and $\hat{\lambda}_1 \geq \cdots \geq \hat{\lambda}_p$ respectively. Fix $1 \leq r \leq s \leq p$ and assume that $\min(\lambda_{r-1} - \lambda_r, \lambda_s - \lambda_{s+1} > 0)$ where $\lambda_0 := \infty$ and $\lambda_{p+1} := -\infty$. Let $d := s - r + 1$, and let $V = (v_r, v_r + 1, \ldots, v_s) \in \mathbb{R}^{p \times d}$ and $\hat{V} = (\hat{v}_r, \hat{v}_{r+1}, \ldots, \hat{v}_s) \in \mathbb{R}^{p \times d}$ have orthogonal columns satisfying $\Sigma v_j = \lambda_j v_j$ and $\hat{\Sigma}\hat{v}_j = \hat{\lambda}_j \hat{v}_j$ for $j = r, r+1, \ldots, s$. Then:

$$\|\sin\Theta(\hat{V}, V)\|_{\mathrm{F}} \leq \frac{2\min(d^{1/2}\|\hat{\Sigma} - \Sigma\|_{\mathrm{op}}, \|\hat{\Sigma} - \Sigma\|_{\mathrm{F}})}{\min(\lambda_{r-1} - \lambda_r, \lambda_s - \lambda_{s+1})}.$$

Moreover, there exists an orthogonal matrix $\hat{O} \in \mathbb{R}^{d \times d}$ such that:

$$\|\hat{V}\hat{O} - V\|_{\mathrm{F}} \leq \frac{2^{3/2}\min(d^{1/2}\|\hat{\Sigma} - \Sigma\|_{\mathrm{op}}, \|\hat{\Sigma} - \Sigma\|_{\mathrm{F}})}{\min(\lambda_{r-1} - \lambda_r, \lambda_s - \lambda_{s+1})}.$$

Here, $\|sin(\Theta(V_1, \hat{V}_1))\|$ is the angle between subspaces $V_1$ and $\hat{V}_1$.

Let $\lambda_i^{(\infty)}$ denote the $i^{\text{th}}$ eigenvalue of $\mathcal{T}_\infty$ and $\lambda_i^{(0)}$ denote the $i^{\text{th}}$ eigenvalue of $\mathcal{T}_0$. Then from Lemma 1, we see that any $\mathcal{T}_\infty$ corresponds to a $\mathcal{T}_0$ with a smaller eigengap between eigenvalues $(\lambda_i - \lambda_{i+1})$ for $i \geq k$ Then, using the theorem above, we can conclude that for $i \leq k$, the upper bound is smaller for $\mathcal{T}_\infty$ than for $\mathcal{T}_0$. Since this upper bound is sharp [21], this proves the theorem.

□

**Remark 1.** Note that Theorem 3 holds true for any $V = \text{span}(\lambda_i, \ldots, \lambda_{i+j})$ for $i + j \leq k, j \geq 0$, that is, the subspace spanned by any range of $(j + 1)$ "large" eigenvalues.

Here, the focus on large eigenvalues is particularly relevant for the problem of community detection. To see this, consider an undirected network with $k$ connected components. Such a network can be represented as a block-diagonal, symmetric DSM with $k$ degenerate eigenvalues equal to 1. In a real-world setting, there may be true edges that violate the block-diagonal structure. If we treat these edges as small perturbations over the block-diagonal matrix, by Weyl's inequality, we expect the eigenvalues of the perturbed matrix to remain close to those of the original (block-diagonal) matrix. i.e., the eigenvalues remain close to 1.

**Remark 2.** Lemma 1 provides insight about the role of $\alpha$. Recall that $c(\alpha)$ is an increasing function of $\alpha$. In our experiments we have used $\alpha = 0.85$ corresponding to $c(\alpha) = 0.78$. Since the eigenvalues are restricted to stay within the $[0, 1]$ interval and to preserve their signs, the algorithm compresses the gap between small eigenvalues (i.e., eigenvalues below $c(\alpha)$) in order to expand the gap between large eigenvalues (i.e., eigenvalues above $c(\alpha)$). We make the following three observations: $\alpha$ controls: (1) which interval will go through compression and which interval will go through expansion, (2) the intensity of this compression/expansion, and (3) the non-linearity of this compression/expansion.

Figure 3C empirically shows that the results of NE are not sensitive to the value of $\alpha$. This stability is due to the relative flatness of $c(\alpha)$, $c(0.15) = 0.6$, $c(0.85) = 0.78$, indicating that the expansion region is not very sensitive to the value of $\alpha$ away from the extreme ends. At the end points of $\alpha$, $c(\alpha)$ changes rapidly. For example, when $\alpha = 1$ the algorithm reduces to a simple diffusion algorithm (without a restart). In that case, the expansion region is only $\{1\}$ and all other eigenvalues are compressed to $\{0\}$ as is expected in a pure diffusion algorithm.

# Supplementary References